\documentclass{emulateapj}
\usepackage{apjfonts}

\newcommand{\Msun}      {\mbox{$\rm\,M_{\mathord\odot}$}}

\begin{document}

\lefthead{Black Hole Accretion Geometry}
\righthead{Tomsick et al.}

\submitted{Accepted by the Astrophysical Journal}

\def\lsim{\mathrel{\lower .85ex\hbox{\rlap{$\sim$}\raise
.95ex\hbox{$<$} }}}
\def\gsim{\mathrel{\lower .80ex\hbox{\rlap{$\sim$}\raise
.90ex\hbox{$>$} }}}

\title{Broadband X-ray Spectra of GX 339--4 and the Geometry of 
Accreting Black Holes in the Hard State}

\author{John A. Tomsick\altaffilmark{1},
Emrah Kalemci\altaffilmark{2},
Philip Kaaret\altaffilmark{3},
Sera Markoff\altaffilmark{4},
Stephane Corbel\altaffilmark{5},
Simone Migliari\altaffilmark{6},
Rob Fender\altaffilmark{7},
Charles D. Bailyn\altaffilmark{8},
Michelle M. Buxton\altaffilmark{8}}

\altaffiltext{1}{Space Sciences Laboratory, 7 Gauss Way, 
University of California, Berkeley, CA 94720-7450, USA
(e-mail: jtomsick@ssl.berkeley.edu)}

\altaffiltext{2}{Sabanci University, Orhanli - Tuzla, Istanbul, 
34956, Turkey}

\altaffiltext{3}{Department of Physics and Astronomy,
University of Iowa, Van Allen Hall, Iowa City, IA 52242, USA.}

\altaffiltext{4}{Astronomical Institute `Anton Pannekoek', 
University of Amsterdam, Kruislaan 403, 1098 SJ, Amsterdam, 
The Netherlands.}

\altaffiltext{5}{AIM - Unit\'e Mixte de Recherche CEA - CNRS -
Universit\'e Paris VII - UMR 7158, CEA-Saclay, Service d'Astrophysique,
91191 Gif-sur-Yvette Cedex, France.}

\altaffiltext{6}{Center for Astrophysics and Space Sciences,
University of California San Diego, 9500 Gilman Dr., La Jolla, 
CA 92093-0424, USA.}

\altaffiltext{7}{School of Physics and Astronomy, University of 
Southampton, Hampshire SO17 1BJ, United Kingdom.}

\altaffiltext{8}{Department of Astronomy, Yale University, 
P.O. Box 208101, New Haven, CT 06520, USA.}

\begin{abstract}

A major question in the study of black hole binaries involves our 
understanding of the accretion geometry when the sources are in 
the ``hard'' state.  In this state, the X-ray energy spectrum is 
dominated by a hard power-law component and radio observations 
indicate the presence of a steady and powerful ``compact'' jet.  
Although the common hard state picture is that the accretion disk
is truncated, perhaps at hundreds of gravitational radii ($R_{\rm g}$)
from the black hole, recent results for the recurrent transient 
GX~339--4 by Miller and co-workers show evidence for optically thick 
material very close to the black hole's innermost stable circular
orbit.  That work focused on an observation of GX~339--4 at a 
luminosity of about 5\% of the Eddington limit ($L_{\rm Edd}$) and 
used parameters from a relativistic reflection model and the 
presence of a soft, thermal component as diagnostics.  In this work, 
we use similar diagnostics, but extend the study to lower luminosities 
(2.3\% and 0.8\% $L_{\rm Edd}$) using {\em Swift} and {\em RXTE} 
observations of GX~339--4.  We detect a thermal component with an 
inner disk temperature of $\sim$0.2 keV at 2.3\% $L_{\rm Edd}$.  At 
0.8\% $L_{\rm Edd}$, the spectrum is consistent with the presence of 
such a component, but the component is not required with high confidence.  
At both luminosities, we detect broad features due to iron K$\alpha$ 
that are likely related to reflection of hard X-rays off the optically 
thick material.  If these features are broadened by relativistic effects, 
they indicate that optically thick material resides within 10 $R_{\rm g}$ 
down to 0.8\% $L_{\rm Edd}$, and the measurements are consistent with 
the inner radius of the disk remaining at $\sim$4 $R_{\rm g}$ down to 
this level.  However, we also discuss an alternative model for the 
broadening, and we note that the evolution of the thermal component is 
not entirely consistent with the constant inner radius interpretation.  
Finally, we discuss the results in terms of recent theoretical work by 
Liu and co-workers on the possibility that material may condense out 
of an Advection-Dominated Accretion Flow to maintain an inner optically 
thick disk.

\end{abstract}


\section{Introduction}

Over the past several years, progress has been made in constraining the geometry 
of black hole accretion disks.  Most binaries with stellar mass black holes are 
X-ray transients, and X-ray observations made when the sources are bright 
($L_{\rm x} \sim 10^{37-39}$ ergs~s$^{-1}$) have uncovered iron K$\alpha$ 
emission lines with broad and redshifted profiles thought to be produced via 
fluorescence when hard X-rays reflect off optically thick disk material 
\citep{rn03}.  Fitting the energy spectra with relativistic reflection models 
yield inner disk radii close to the innermost stable circular orbit (ISCO) of 
the black hole \citep{miller02}.  Another constraint comes from the high 
frequency (100--500 Hz) quasi-periodic oscillations (QPOs) that are present 
for a number of black hole systems \citep{remillard02}.  Although the origin 
of these QPOs is still unclear, if they are related to Keplerian orbital 
frequencies for matter in an optically thick \cite[e.g.,][]{ss73} accretion 
disk, they correspond to time scales very close to the ISCO, implying that 
the disk must extend near the black hole.  Furthermore, these constraints on 
the inner radius of the accretion disk ($R_{\rm in}$) occurred at times when the 
X-ray energy spectrum included a strong thermal component that is consistent 
with the \cite{ss73} model with an inner disk temperature of $\sim$1 keV and 
an inner radius close to or at the ISCO.

As black hole outbursts evolve, they enter different spectral states, and one 
definition for the various states was recently described in \cite{mr06}.  While the 
iron lines, QPOs, and thermal components discussed above have provided a determination
of $R_{\rm in}$ in the ``thermal-dominant'' and ``steep power-law'' (SPL) spectral 
states, in this work, we focus on the hard state where the accretion disk 
geometry is still unclear.  The hard state is most often observed at the 
beginning and end of the outburst and is characterized by a hard energy spectrum 
and a high level of X-ray timing noise.  At the ends of outbursts, the hard state
is most often seen when sources reach Eddington-scaled luminosities below 
$L/L_{\rm Edd} = 0.01$--0.04 \citep{maccarone03}, but the hard state also can
occur at higher luminosities, especially when outbursts commence.  Radio 
observations have made it clear that a steady outflow in the form of a ``compact'' 
jet is characteristic of the hard state \citep{fender01}, and, at least in some 
systems, the jet power is inferred to be more than 20\% of the X-ray luminosity 
\citep{fender01b}.  Thus, one reason that it is important to understand the disk 
geometry and other physical processes that occur in the hard state is that this 
is the one state that is linked to the production of a steady and powerful jet.

The common picture for the hard state accretion disk has been that its inner 
edge recedes (i.e., $R_{\rm in}$ increases), leaving a hot flow, such as an 
Advection-Dominated Accretion Flow (ADAF) or a spherical corona, where most 
of the X-ray emission is produced via Comptonization \citep{nmy96,emn97}.  
Indeed, it has been shown that evaporation of the accretion disk can lead to 
a truncated disk at low mass accretion rates \citep{mlm00}.  From an observational 
standpoint, several measurements are suggestive of a truncated disk, but no one 
method provides a direct measurement of $R_{\rm in}$.  Examples of measurements 
that support an increase in $R_{\rm in}$ include a rapid drop in the temperature 
and flux of the thermal component (often to non-detection in the $>$3 keV band) 
as the source enters the hard state as well as a drop in the characteristic 
frequencies seen in the power spectra \citep{pkr97,dove97,tk00,rgc01,kalemci04,tkk04}.  
In addition, a gradual drop in the strength of the reflection component can be 
explained if the disk is truncated and there is overlap between the disk and corona 
\citep{zls99,zdziarski03}.  

While these measurements can be explained within the truncated disk picture, other 
explanations have also been suggested.  Much of the evidence for the disappearance 
of the thermal component come from {\em RXTE} spectra that lack soft X-ray coverage, 
allowing for the possibility that the thermal component simply shifts out of the 
instrumental bandpass.  Also, an increase in the ionization state of the inner disk 
could cause a drop in the strength of the reflection component.
In addition, some recent X-ray observations of black holes in their hard state
cast doubt on whether the truncated disk picture is correct.  The black hole 
system GX~339--4 was observed in the hard state with {\em XMM-Newton} and the 
{\em Rossi X-ray Timing Explorer (RXTE)}, and a fit with reflection and
iron line emission components accounting for relativistic smearing effects
are consistent with an inner disk radius very close to the ISCO \citep{miller06a}.
The ISCO is at $6 R_{\rm g}$ (where $R_{\rm g} = GM/c^{2}$, $G$ and $c$ are 
constants, and $M$ is the black hole mass) for a non-rotating black hole and
at $1.23 R_{\rm g}$ for a black hole rotating at the limiting spin rate of 
$a_{*} = 0.9982$ \citep{thorne74}, where $a_{*}$ is the mass-normalized angular 
momentum parameter.  \cite{miller06a} obtained a value of $R_{\rm in}/R_{\rm g} 
= 4.0\pm 0.5$ from the GX~339--4 hard state X-ray spectra.  As this value is only 
slightly larger than the value of 2--3 $R_{\rm g}$ that \cite{miller04} obtained 
for GX~339--4 in the SPL state, it suggests that there is very little change in 
$R_{\rm in}$ between the two states.  However, we must keep in mind that the 
interpretation of the relativistic smearing parameters depend on the model being 
physically correct, and other explanations have been advanced for iron line 
broadening \citep[e.g.,][]{lt07}.

For the thermal component from an optically thick disk, one expects the emission 
to fall outside the X-ray regime if $R_{\rm in}$ increases dramatically (e.g., to 
hundreds of $R_{\rm g}$ or more) as predicted by ADAF models \citep{mcclintock03}.  
However, there are numerous examples of black hole systems, including well-known 
systems such as Cygnus~X-1 and GX~339--4, that exhibit thermal X-ray emission 
with temperatures 0.1--0.4 keV when they are in the hard state
\citep{ebisawa96,zdziarski98}.  While this implies that very large values of 
$R_{\rm in}$ are not required for black holes to enter the hard state, the 
thermal components have been found mostly only when the systems are in the 
brightest phases of their hard states.  For example, the 
{\em XMM-Newton}/{\em RXTE} observation of GX~339--4 in the hard state described 
above occurred at the start of an outburst at $L/L_{\rm Edd} = 0.05$ and spectral 
fits yielded a thermal component with a temperature of $\sim$0.4 keV.  
X-ray observations of other black hole systems at lower X-ray luminosities 
have provided relatively high quality X-ray spectra, and, in some cases, soft 
components have been detected \citep[e.g.,][]{miller06b}, while, in other cases, 
they have not \citep[e.g.,][]{mcclintock01,tkk04}.

For this paper, we have obtained broadband {\em Swift} \citep{gehrels04} and 
{\em RXTE} \citep{brs93} observations of the recurrent transient GX~339--4 in 
the hard state to constrain the accretion disk geometry in this state.  The 
observations were made at the end of the most recent outburst that began in 
2007 January \citep{miller07} after the transition to the hard state that 
occurred in 2007 May \citep{kalemci07}.  The luminosities of the observations we 
use in this work are well below $L/L_{\rm Edd} = 0.05$, and our goal is to use 
the spectra along with reflection models that account for relativistic 
effects to constrain the accretion geometry at these low luminosities.

\section{Observations}

\subsection{{\em RXTE} Monitoring Observations}

We obtained daily pointed {\em RXTE} monitoring observations of GX~339--4
starting on 2007 April 20 (MJD 54,210) when the flux dropped below
$4\times 10^{-9}$ ergs~cm$^{-2}$~s$^{-1}$ as measured in the 1.5--12~keV
band by the {\em RXTE} All-Sky Monitor.  Observations were made under
our program (proposal \#92704) until 2007 July 18 (MJD 54,299), and the
{\em RXTE} Proportional Counter Array \citep[PCA][]{jahoda06} 3--25~keV 
light curve during
this 89 day period is shown in Figure~\ref{fig:lc}.  The exposure times
were typically 1--3~ks for these monitoring observations although some 
longer {\em RXTE} observations were also obtained as described below.
We reported an increase in timing noise and a hardening of the GX~339--4 
energy spectrum that indicates a change to the hard-intermediate state 
\cite[see][]{hb05} occurred on 2007 May 12 \citep{kalemci07}, and this 
is marked in Figure~\ref{fig:lc} with a vertical dotted line at MJD 54,232.  
By 2007 May 22 (marked in Figure~\ref{fig:lc} by the vertical dashed line 
at MJD 54,242), the source had reached the hard state with an energy 
spectrum dominated by a power-law with a photon index of $\Gamma = 1.6$.
Further evidence that the source reached the hard state by this time
includes an increase in the optical and infrared flux \citep{bb07}
as well as core radio flux (S. Corbel, private communication), which 
is an indication for the presence of a compact jet.

\subsection{{\em Swift} and {\em RXTE} Observations}

\begin{figure}
\plotone{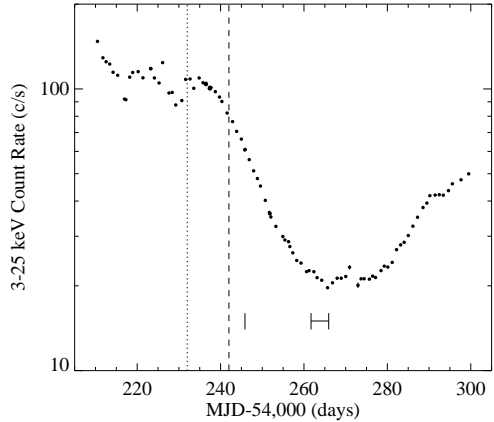}
\vspace{0.0cm}
\caption{The 3--25 keV light curve for GX~339--4 as measured by the
{\em RXTE}/PCA.  The rates shown are for a single Proportional 
Counter Unit (PCU2).  The vertical dotted line shows the start of
the transition to the hard state, and the vertical dashed line shows
when the source reached the hard state.  The times of the {\em Swift}
observations at MJD 54,245 and MJD 54,261--54,265 are marked.\label{fig:lc}}
\end{figure}

The main focus of this work is the study of the broadband $\sim$1--200 keV
energy spectra from GX~339--4 at two times after the source made a 
transition to the hard state.  As indicated in Figure~\ref{fig:lc}, 
the first {\em Swift} observation occurred on 2007 May 25 (MJD 54,245),
a few days after the transition to the hard state.  The average {\em Swift}
X-ray Telescope \citep[XRT][]{burrows05}
count rate during the 6150~s observation was 
$10.41\pm 0.04$ c~s$^{-1}$ (0.8--8 keV).  As indicated in Table~\ref{tab:obs},
we obtained {\em RXTE} observations that were simultaneous with
much of the {\em Swift} observation.  The average PCA count rate was 
$44.1\pm 0.1$ c~s$^{-1}$ (3.6--25 keV) for Proportional Counter Unit (PCU) 
\#2, and the average High Energy X-ray Timing Experiment 
\citep[HEXTE][]{rothschild98} count rate 
was $10.4\pm 0.2$ c~s$^{-1}$ (17--240 keV) for HEXTE cluster B.  As 
described below, Spectrum \#1 consists of the spectra from XRT, PCA, 
and HEXTE from these observations.

We used the same 3 instruments for Spectrum \#2, but the observations
were made 2--3 weeks later during the time period 2007 June 10--14.  
Table~\ref{tab:obs} shows that there were 3 {\em Swift} and 5 {\em RXTE}
observations made during this time frame.  Although the {\em Swift}
and {\em RXTE} observations were mostly not strictly simultaneous, 
the source did not show large changes in count rate or spectral hardness
over the 5 day period.  The 3.6--25 keV PCA count rates for each of the
5 observations listed in Table~\ref{tab:obs} were $16.4\pm 0.1$ c~s$^{-1}$,
$16.2\pm 0.1$ c~s$^{-1}$, $15.7\pm 0.2$ c~s$^{-1}$, $15.1\pm 0.1$ c~s$^{-1}$, 
and $14.2\pm 0.1$ c~s$^{-1}$, showing a gradual but not dramatic decline.  
As shown in Figure~\ref{fig:lc}, this period is close to the minimum flux 
that GX~339--4 obtained before its flux began increasing again.  The average 
{\em Swift} XRT count rate was $2.86\pm 0.02$ c~s$^{-1}$ (0.8--8 keV), while 
the average PCU2 and HEXTE-B count rates were $15.6\pm 0.1$ c~s$^{-1}$ 
(3.6--25 keV) and $3.8\pm 0.2$ c~s$^{-1}$ (17--240 keV), respectively.  
Thus, when compared to Spectrum \#1, the XRT count rate is lower by a factor 
of 3.6 while the count rates for the {\em RXTE} instruments are lower by a 
factor of $\sim$2.8.

\section{Spectral Analysis}

We performed spectral analysis using XRT, PCA, and HEXTE data from the
observations described above.  To extract the spectra, we used the 
{\em Swift} and {\em RXTE} tools provided in the HEASOFT v6.3.1 software
package.  The XRT instrument consists of a CCD imager at the focus of
a grazing incidence X-ray telescope.  Although the instrument is capable
of two-dimensional imaging for faint sources, our target was bright 
enough to require the use of Windowed Timing mode, which provides
one-dimensional imaging.  For the GX~339--4 energy spectra, we determined 
the one-dimensional centroid of the target and extracted the photons
within $47^{\prime\prime}$ of the centroid.  We also extracted background 
spectra that include photons from regions between $77^{\prime\prime}$
and $171^{\prime\prime}$ from the GX~339--4 centroid.  We used version 9 
of the XRT response matrix and kept events with grades in the range 0--2 
({\ttfamily swxwt0\_20010101v009.rmf}).  In addition, we produced an 
ancillary response file (ARF) with the HEASOFT tool {\ttfamily xrtmkarf 
v0.5.3}.  In making the ARF file, we used an exposure map produced with 
the tool {\ttfamily xrtexpomap v0.2.2}.

For the PCA spectra, we used the {\ttfamily SkyVLE} background model 
that was most recently updated 2005 November 28\footnote{A problem with 
the software that produces PCA background spectra was recently 
announced (see http://heasarc.gsfc.nasa.gov/docs/xte/xte\_1st.html).  
We re-extracted the spectra with the corrected software and found that
the problem did not impact our observations.}, and we used the HEASOFT
tool {\ttfamily pcarsp v10.1} to produce the response matrix.  Although 
the PCA has 5 PCUs, it is typical that only 2 or 3 PCUs will be turned 
on for a given observation.  In our case, only PCUs 0 and 2 were turned 
on for all of our {\em RXTE} observations.  Due to the loss of the PCU 0
propane layer, this PCU's response is not as well known as the other
PCUs \citep{jahoda06}.  Also, for all the PCUs, the response for the
top anode layer (where most of the counts are detected) is better modeled 
than the bottom layers.  Thus, for this study, we only used the top
anode layer of PCU 2 when extracting spectra.  To check on the level
of systematic error, we extracted the PCU 2 spectrum for the Crab 
nebula using Observation ID 92802-01-22-00, which is an observation 
from 2007 May 14.  Fitting the spectrum with an absorbed power-law, 
we find that 1\% systematic errors are required to reach a 
reduced-$\chi^{2}$ near 1.0, and we use 1\% systematics for the GX~339--4
spectra as well.  For the HEXTE spectra, we used only HEXTE-B because 
HEXTE-A no longer obtains background measurements.

We performed preliminary spectral fits to Spectrum \#1 to check on the
agreement between the calibrations of the different instruments.  For
these preliminary fits, we used XRT, PCA, and HEXTE data in the 
0.3--10 keV, 2.75--25 keV, and 17--240 keV energy bands, respectively.
We used XSPEC v12 and fitted the spectra with an absorbed power-law
model.  We also included a multiplicative constant in the model to
allow for differences in the overall normalizations between instruments.
The fact that we obtain a poor fit ($\chi^{2}/\nu = 1282/269$) is 
partially due to the fact that the power-law model is too simple
as well as being due to instrument calibrations.
For example, it is known that the XRT calibration is complicated
by features due to the SiO$_{2}$ layer in the CCD detectors at
low energies \citep{osborne05}.  In fact, we see a sharp dip in the 
XRT residuals near 0.5--0.6 keV that is likely instrumental.  
Furthermore, we see that the XRT calibration does not agree with 
the PCA calibration above 8 keV, and these issues lead us to use 
the XRT in the 0.8--8 keV energy range for the fits described below.  
In addition, the PCA calibration does not agree with XRT at the very 
bottom of the PCA range as there are strong positive residuals in 
the PCA spectrum below 3.6 keV.  Thus, in the following, we use the 
3.6--25 keV PCA spectrum.  PCA and HEXTE match well in the region 
where they overlap, and we use the full 17--240 keV HEXTE bandpass.  
In addition, we note that there is good agreement (within 5\%) between 
the overall normalizations of the three instruments.

For PCA, Spectra \#1 and \#2 have high statistical quality (hundreds
or thousands of counts per energy bin) across the bandpass, but some
rebinning was required for XRT and HEXTE.  For XRT, we rebinned from
719 channels to 148 channels, leaving averages of 432 counts/bin
and 259 counts/bin for Spectra \#1 and \#2, respectively.  We 
rebinned the HEXTE spectra from 210 channels to 22 channels.

\section{Results}

\subsection{Fits with Basic Models:  Is a Thermal Disk Component Present?}

\begin{figure}
\centerline{\includegraphics[width=0.45\textwidth]{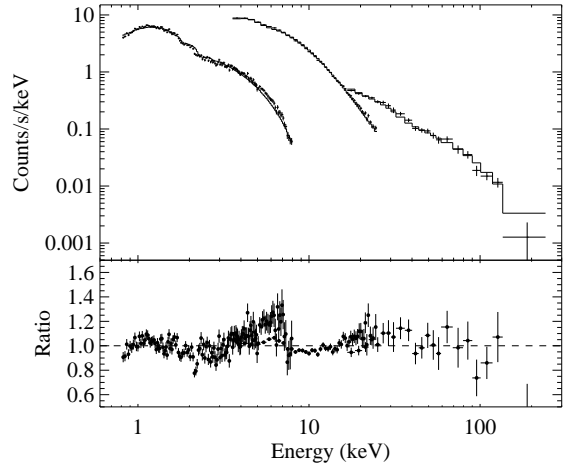}}
\vspace{-1.2cm}
\caption{Spectrum \#1, including data from {\em Swift}/XRT (0.8--8 keV), 
{\em RXTE}/PCA (3.6--25 keV), and {\em RXTE}/HEXTE (17--240 keV), fitted 
with an absorbed power-law model.  The data-to-model ratio is shown in 
the bottom panel.\label{fig:spectrum1_pl}}
\end{figure}

\begin{figure}
\centerline{\includegraphics[width=0.45\textwidth]{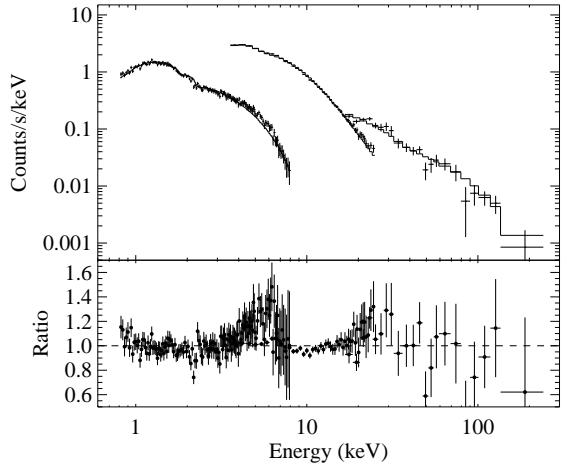}}
\vspace{-1.2cm}
\caption{Spectrum \#2, including data from the XRT, PCA, and HEXTE 
instruments, fitted with an absorbed power-law model.  The data-to-model 
ratio is shown in the bottom panel.\label{fig:spectrum2_pl}}
\end{figure}

We fitted the spectra with an absorbed power-law model.  To account 
for absorption, we use the photoelectric absorption cross sections 
from \cite{bm92} and elemental abundances from \cite{wam00}, which 
correspond to the estimated abundances for the interstellar medium.  
A power-law model does not provide a good fit to either spectrum, with 
$\chi^{2}$ equal to 651.8 and 321.6 for Spectra \#1 and \#2, respectively 
(for 213 degrees of freedom in both cases).  Figures~\ref{fig:spectrum1_pl} 
and \ref{fig:spectrum2_pl} show the spectra and the residuals in the
form of a data-to-model ratio.  In both cases, the strongest feature
in the residuals occurs in the region near the iron K$\alpha$ region.
Positive residuals are present close to the $\sim$6--7 keV range where
an emission line might be expected to be present, and negative residuals
are seen from 7 keV to beyond 10 keV.  The positive residuals that are
present in the 20--40 keV range, especially in Spectrum \#1, may be
due to a reflection component.  Finally, the curvature in the residuals
at the lower end of Spectrum \#1 may indicate the presence of a 
thermal component from an optically thick accretion disk.  The fit
parameters are given in Table~\ref{tab:basic}, and it is notable 
that column density of $N_{\rm H} = (3.1\pm 0.1)\times 10^{21}$ cm$^{-2}$
obtained for the power-law fit to Spectrum \#1 is somewhat lower than
the value inferred from the work of \cite{hynes04}.  From optical
observations, \cite{hynes04} prefer a value of $E(B-V)\gsim 0.85$, which
corresponds to $N_{\rm H}\gsim 4.7\times 10^{21}$ cm$^{-2}$ using
conversions given in \cite{ps95}.  Thus, the lower column density 
could also be an indication that a thermal component is present.  

To test whether a thermal component is present in the spectra, we performed 
the fits detailed in Table~\ref{tab:basic}.  For Spectrum \#1, when we add a 
disk-blackbody \citep{mitsuda84} component to the power-law, the quality of 
the fit shows a large improvement to $\chi^{2}/\nu = 532.0/211$.  This, in 
addition to the fact that the $N_{\rm H}$ increases to a level which is 
consistent with the lower limit from optical extinction measurements, are 
indications for the presence of a thermal component.  However, even with the 
disk-blackbody component, the fit is poor due to the iron features in the 
spectrum, so we re-fitted Spectrum\#1 with a model that takes these features 
into account.  Although we use a more physical model for the iron line and 
reflection below, here we add a smeared iron edge \citep{ebisawa94} to the 
model because it is a simple addition that significantly improves the fit.  
When the disk-blackbody component is added to a model with a power-law and 
a smeared edge, the fit shows a large improvement from $\chi^{2}/\nu = 
444.7/211$ to 338.3/209.  

We carried out the same series of fits for Spectrum \#2, and while they 
also provide evidence for the presence of a thermal component, the evidence 
is considerably weaker than for Spectrum \#1.  For the models without the 
smeared edge (see Table~\ref{tab:basic}), adding the disk-blackbody
component gives a relatively large improvement in the fit from $\chi^{2}/\nu = 
321.6/213$ to 280.1/211.  However, with the smeared edge, the change from
$\chi^{2}/\nu = 210.6/211$ to 200.2/209 is rather small.

\begin{figure}
\plotone{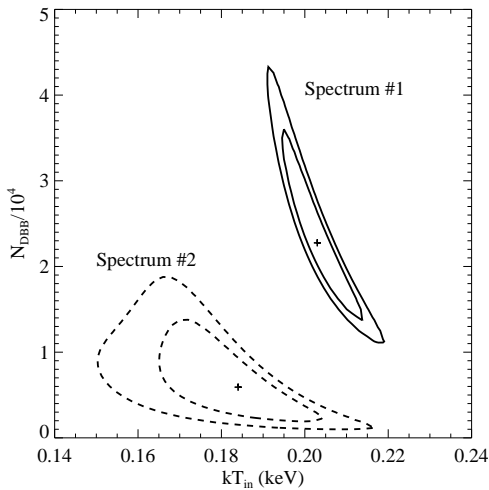}
\vspace{0.0cm}
\caption{Confidence contours for the thermal (disk-blackbody) components for 
Spectra \#1 (solid contours) and \#2 (dashed contours).  In each case, the 
inner-most contour encircles the 68\% confidence ($\Delta\chi^{2} = 2.30$)
error region for the two parameters and the outer-most contour corresponds
to 90\% confidence ($\Delta\chi^{2} = 4.61$).\label{fig:contour}}
\end{figure}

To further test the significance of the thermal component in the spectra
and to determine if there is any evolution in the thermal component between
Spectra \#1 and \#2, we produced error contours for the disk-blackbody
temperature ($kT_{\rm in}$) and normalization ($N_{\rm DBB}$).  In 
Figure~\ref{fig:contour}, the outermost contour for each spectrum corresponds
to 90\% confidence for two-parameters of interest ($\Delta\chi^{2} = 4.61$).
The error region for Spectrum \#1 shows that $N_{\rm DBB}$ is significantly
different from zero, consistent with the presence of a thermal component
in Spectrum \#1.  The error region for Spectrum \#2 is well-separated from
that of Spectrum \#1, showing a clear change in the thermal component 
between the two spectra.  The results indicate a drop in $kT_{\rm in}$
or $N_{\rm DBB}$ or both parameters.  To estimate the significance of
the thermal component in Spectrum \#2, we adjusted $\Delta\chi^{2}$ 
until the confidence contour reached a $N_{\rm DBB}$ value of zero.
The 99\% confidence contour ($\Delta\chi^{2} = 9.21$) does not reach
zero, but zero is reached with a slightly larger contour
($\Delta\chi^{2} = 10$), which is consistent with a $\sim$1\% chance 
that this component is spurious.

\subsection{Iron Line and Reflection Modeling}

We re-fitted the spectra in an attempt to improve our modeling of 
the iron line and reflection.  Initially, we removed the smeared edge
and fitted the spectra with a model consisting of a disk-blackbody, 
a power-law, and a Gaussian emission line.  Although the Gaussian
greatly improves the fit in both cases, the energy is well below
the 6.4--7.1 keV iron range with values of $E_{\rm line} = 
3.8^{+0.6}_{-0.8}$ keV and $4.0^{+0.8}_{-1.4}$ keV for Spectra
\#1 and \#2, respectively.  As suggested by the residuals in 
Figures~\ref{fig:spectrum1_pl} and \ref{fig:spectrum2_pl}, the 
Gaussians are also very broad with widths of $\sigma\sim 2.0$ keV
in both cases.  This suggests that we may be seeing relativistically
smeared iron lines.

The {\ttfamily laor} emission line model \citep{laor91} is 
appropriate for reflection from an accretion disk around a rotating 
black hole, and if the iron line is due to reflection, then one 
expects to see evidence for excess emission from reflection at 
higher energies ($\sim$20--40 keV) as well.  In fact, the positive 
residuals for Spectrum \#1 (and to some extent for Spectrum \#2) in 
this energy range suggest that there is indeed a reflection 
component.  Thus, we re-fitted the spectra with a model consisting 
of a disk-blackbody, an iron emission line, and a {\ttfamily pexriv}
reflection component \citep{mz95}, which includes both direct
emission from a power-law as well as emission reflected off an
accretion disk with neutral or partially ionized material.  While 
the {\ttfamily pexriv} model includes the ionization effects, which 
are likely important for the hot accretion disks around X-ray binaries, 
it does not include relativistic smearing.  Thus, in our model, we 
convolved the {\ttfamily pexriv} model with the {\ttfamily laor} 
model shape using the XSPEC convolution model {\ttfamily kdblur}.  
We set up the XSPEC model so that {\ttfamily kdblur} convolves
the sum of a narrow iron line and the {\ttfamily pexriv} model.
The {\ttfamily kdblur} parameters include $R_{\rm in}$ (the inner 
radius of the disk), $R_{\rm out}$ (the outer radius of the disk), 
$i$ (the binary inclination), and $q$ (the power-law index for the 
radial emissivity profile).

\begin{figure}
\plotone{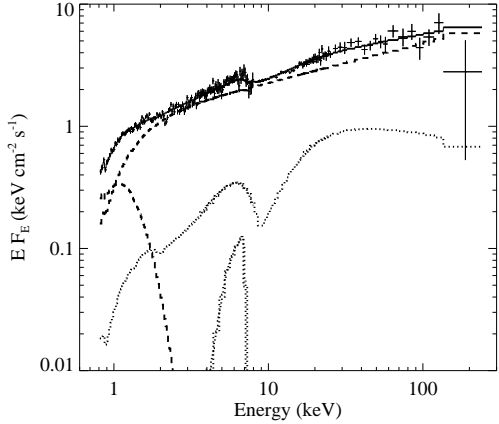}
\vspace{0.0cm}
\caption{Spectrum \#1 fitted with the model detailed in Table~\ref{tab:reflection}
and plotted in flux units.  The various model components are shown and include
a thermal disk-blackbody component, a power-law, a reflection component, and
an iron emission line (the last two include relativistic effects).\label{fig:efe1}}
\end{figure}

\begin{figure}
\plotone{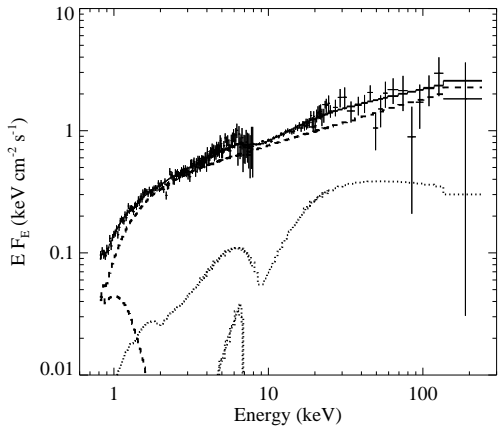}
\vspace{0.0cm}
\caption{Spectrum \#2 fitted with the model detailed in Table~\ref{tab:reflection}
and plotted in flux units.  The various model components are shown and include
a thermal disk-blackbody component, a power-law, a reflection component, and
an iron emission line (the last two include relativistic effects).\label{fig:efe2}}
\end{figure}

The results of these fits are given in Table~\ref{tab:reflection},
and the fitted spectra are shown in Figures~\ref{fig:efe1} and 
\ref{fig:efe2}.  The power-law and disk-blackbody parameters 
have similar values to those that we found with the basic models
(Table~\ref{tab:basic}), and the quality of the fits is better 
than we obtained with the basic models.  For both spectra, the
presence of the {\ttfamily pexriv} reflection component is 
required at high confidence as indicated by the fact that the
reflection covering fraction ($\Omega/2\pi$) is significantly
different from zero.  The exact value of $\Omega/2\pi$ depends
strongly on the binary inclination, which is not known for 
GX~339--4.  Previous fits to higher quality X-ray spectra have 
given a value of $i = 20^{+5}_{-15}$ degrees \citep{miller06a},
and we adopt a value of $20^{\circ}$ to facilitate comparisons to 
the previous work.  With this inclination, we obtain values of
$\Omega/2\pi = 0.22^{+0.06}_{-0.05}$ and $0.24^{+0.11}_{-0.08}$
for Spectra \#1 and \#2, respectively.  The {\ttfamily pexriv} 
ionization parameter, $\xi$, is not very well-constrained, but 
it is significantly greater than zero for both spectra, indicating 
a disk that is at least partially ionized.

While the reflection component is statistically significant
in both Spectra \#1 and \#2, an additional iron line in the
6.4--7.1 keV range (see $E_{\rm line}$ in Table~\ref{tab:reflection})
is required for Spectrum \#1 but is required at only slightly
more than 90\% confidence for Spectrum \#2 as shown by the values 
of the emission line normalization ($N_{\rm line}$) given in 
Table~\ref{tab:reflection}.  This is not due to the lack of iron 
features in Spectrum \#2 but because, with the relativistic 
broadening, the reflection component contains a bump related to 
the iron absorption edge that can mimic a broad iron emission 
line (see Figures~\ref{fig:efe1} and \ref{fig:efe2}).  However, 
even though the emission line is not clearly detected in 
Spectrum \#2, the parameters that account for the relativistic 
smearing are still well-constrained for both spectra because 
both the line and the reflection ({\ttfamily pexriv}) components 
are smeared.

While there are four relativistic smearing parameters, only two
of the parameters are left as free parameters in our fits.  
As mentioned above, we fixed the binary inclination to $20^{\circ}$, 
and we fixed the outer disk radius to $R_{\rm out} = 400 R_{\rm g} 
(= GM/c^{2}$, where $G$ and $c$ are constants and $M$ is the black 
hole mass).  One of the free parameters is the inner disk radius, 
and we find that $R_{\rm in} = 3.6^{+1.4}_{-1.0} R_{g}$ and 
$2.9^{+2.1}_{-0.7} R_{g}$ for Spectra \#1 and \#2, respectively,
implying that the reflecting material is very close to the 
black hole.  The other free parameter is the power-law index for
the radial emissivity profile, and we obtain values consistent
with $q = 3$ for both spectra.

\section{Discussion}

\subsection{Constraints from the Reflection Model}

Our results on the iron line and reflection component of GX~339--4 
join a relatively small number of observations where reflection models 
that account for the relativistic effects near black holes have been 
fitted to broadband X-ray spectra of black holes in the hard state.  
Our findings are most directly comparable to the results of 
\cite{miller06a} where both the GX~339--4 and Cygnus~X-1 showed
evidence for a broad iron K$\alpha$ emission line and a smeared
reflection component while they were in the hard state. 
Table~\ref{tab:reflection} compares the parameters for GX~339--4
from the fits to the 2004 {\em XMM-Newton}/{\em RXTE} spectrum
reported in \cite{miller06a} to the parameters we obtain by fitting 
the same model to our {\em Swift}/{\em RXTE} spectra.  Independent 
of assumptions about the distance to the source, Spectra \#1 and \#2 
were taken when the GX~339--4 luminosity (1--100 keV, unabsorbed) 
was, respectively, 2.4 and 7.0 times lower than the 2004 spectrum.
Adopting a source distance of 8 kpc \citep{hynes04} and a black hole 
mass of 5.8\Msun~\citep{hynes03}, which are the same values used by 
\cite{miller06a}, we estimate that the Eddington-scaled luminosities 
during our observations are $L/L_{\rm Edd} = 0.023$ and 0.008.

The two most significant reflection parameters for answering the 
question of the accretion geometry for GX~339--4 in the hard state 
are the covering fraction $\Omega/2\pi$ and the inner radius of 
the optically thick disk ($r_{\rm in} = R_{\rm in}/R_{\rm g}$).  
The covering fraction of $\sim$0.23 that we obtain (assuming 
$i = 20^{\circ}$) is consistent with the value of $0.22\pm 0.06$ 
obtained by \cite{miller06a}, and this may indicate little change 
in the system geometry even though we are observing at a lower flux.  
However, these values of $\Omega/2\pi$ are much less than the values
of unity or even larger that have been seen for GX~339--4 as well
as other black hole systems in the SPL state \citep{zdziarski03,miller04}.
Although we are including relativistic effects that were not included
in earlier work by \cite{zdziarski03}, our covering fraction and
power-law index ($\Gamma$) values are very similar to the values
\cite{zdziarski03} used to demonstrate a systematic drop in the
covering fraction as the source hardens.  Thus, these results are
consistent with a significant change in geometry between the SPL 
and the hard state, but not in the hard state between 0.056 and 
0.008 $L_{\rm Edd}$.  In addition, we note that geometry is not 
necessarily the only change that occurs between the SPL and the 
hard state.

The inner disk radii implied by the reflection model parameters 
also suggest little or no change in $r_{\rm in}$ down to 0.008 
$L_{\rm Edd}$.  While \cite{miller06a} obtain $r_{\rm in} = 
4.0\pm 0.5$, our reflection fits indicate {\em maximum} (90\% 
confidence) $r_{\rm in}$ values of 5.0 for for Spectra \#1 and \#2, 
and the spectra are consistent with little change from 
$r_{\rm in}\sim 4$ over the 0.008--0.056 $L_{\rm Edd}$ luminosity 
range.  However, for Spectrum \#2, since the iron line is not 
required at high statistical confidence, we re-fitted the 
spectrum without the emission line, allowing only the blurred 
reflection component to constrain $r_{\rm in}$.  In this case, 
we obtain $r_{\rm in} = 4.1^{+5.9}_{-1.4}$ so that we should not 
rule out the possibility that the inner disk radius increases to 
$\sim$10 $R_{\rm g}$ at the lowest luminosity that we are sampling.

Of course, these radius constraints are only valid if the blurring
that is clearly present is caused by the Doppler boosting due to 
the motions of the material around the black hole and the 
gravitational redshift, and it is worthwhile to consider whether 
other physical effects could cause similar blurring.  Probably
the best developed competing model attributes the reflection 
component to Compton downscattering of the source's X-ray emission 
in a large-scale and powerful wind \citep{ts05}.  The model 
requires hard X-ray emission with a power-law photon index of 
$\Gamma < 2$, and the model has been shown to be able to reproduce
the shape of the $\sim$20--200~keV hard state spectra from GX~339--4 
\citep{ts05}.  More recently, it was shown that this Comptonization
model can produce blurred and redshifted iron lines due to 
fluorescence of the wind material \citep{lt07}.  Although this
full model is not currently available for fitting our GX~339--4
spectra, in the future, it would be interesting to see if it
can explain all the spectral features we observe.

\subsection{The Thermal Component in the Hard State}

Further evidence for an optically thick disk in the hard state
that is not highly truncated comes from the presence of a thermal
component that has now been seen in the hard state spectra of several 
black hole systems.  In our case, Spectrum \#1 has a significant 
thermal component while the thermal component may be present in 
Spectrum \#2, but it is not required at very high significance.  
Thus, we focus on comparing our Spectrum \#1 parameters to the 
disk-blackbody parameters found by \cite{miller06a}.  As shown in 
Table~\ref{tab:reflection}, the inner disk temperature of 
$0.193\pm 0.012$ keV for Spectrum \#1 is significantly lower than 
the value of $kT_{\rm in} = 0.39\pm 0.04$ keV measured when the 
3--100 keV flux of the source was 2.4 times brighter.  While a drop 
in temperature does not necessarily signal a change in accretion 
geometry, the disk-blackbody normalization is related to the disk 
inner radius according to $N_{\rm DBB}\propto R_{\rm in}^{2}$, so 
the fact that $N_{\rm DBB}$ is significantly larger for Spectrum \#1 
(see Table~\ref{tab:reflection}) could indicate an increase in 
$R_{\rm in}$.

To further investigate the difference between the values of
$N_{\rm DBB}$ in the two spectra, we note that for these low 
temperatures, the disk component is strongly impacted by interstellar
absorption, and the value of $N_{\rm H}$ that we obtain is somewhat
higher than the value derived by \cite{miller06a}.  Thus, we
refitted Spectrum \#1 after fixing the column density to the
\cite{miller06a} value.  We obtain only a slightly higher 
temperature of $kT_{\rm in} = 0.220\pm 0.013$ keV and 
$N_{\rm DBB} = 6100^{+2800}_{-1500}$.  Thus, $N_{\rm DBB}$ is 
still nearly an order of magnitude higher for Spectrum \#1 compared 
to the value of $\sim$700 obtained by \cite{miller06a}.  Another 
(not independent) way to examine the question of whether $R_{\rm in}$ 
changes is that the disk-blackbody flux ($F_{\rm disk}$) should be 
proportional to $kT_{\rm in}^{4}$ for constant $R_{\rm in}$ 
\citep{mitsuda84}.  Based on the parameters with $N_{\rm H}$ fixed, 
the bolometric disk-blackbody flux for Spectrum \#1 is 
$3.2\times 10^{-10}$ ergs~cm$^{-2}$~s$^{-1}$, while the flux from 
the \cite{miller06a} parameters is $3.4\times 10^{-10}$ 
ergs~cm$^{-2}$~s$^{-1}$.  While the fact that these fluxes are 
nearly the same despite a drop in $kT_{\rm in}$ from 0.39 keV to 
0.22 keV could be explained by a change in $R_{\rm in}$, other 
physical effects could also be important such as possible changes 
in the ``color correction factor'' \citep{st95} or changes in the 
temperature profile in the disk.  It is also worth noting that the 
observations were made with different soft X-ray instruments 
({\em XMM-Newton} vs.~{\em Swift}) and differences in calibration 
could be important.  However, despite these other possibilities, if 
the correct explanation is a change in the inner radius, $R_{\rm in}$ 
would need to change by a factor of $\sim$3.

Our results for the evolution of the disk-blackbody parameters
appear to be in contrast to results obtained recently by 
\cite{rykoff07} for another accreting black hole (XTE~J1817--330) 
in the hard state.  Using $\sim$20 {\em Swift} observations
covering inner disk temperatures from 0.2 to 0.8~keV, 
\cite{rykoff07} found a relationship close to $F_{\rm disk}\propto
kT_{\rm in}^{4}$ (they actually found an index of $3.3\pm 0.1$ or 
$4.3\pm 0.1$ depending on the model they used for the non-thermal 
component).  Although XTE~J1817--330 does show some deviations
from the trend for individual data points 
\citep[see Figure 3 from][]{rykoff07}, none of the deviations from 
the $F_{\rm disk}\propto kT_{\rm in}^{4}$ trend are as large as we 
see for GX~339--4, suggesting that we may be seeing a different 
evolution for GX~339--4.

\subsection{Implications for the Hard State Geometry}

Our spectra of GX~339--4 provide evidence for an optically thick 
accretion disk in the hard state.  At $L/L_{\rm Edd} = 0.023$, the 
evidence comes from both blurred reflection and iron line features 
as well as a significant thermal component.  At $L/L_{\rm Edd} = 
0.008$, the evidence primarily comes from the blurred reflection 
component.  The most constraining measurement of $r_{\rm in}$ comes 
from the reflection component, and these measurements require 
$r_{\rm in} < 5$ at 0.023 $L_{\rm Edd}$ and $r_{\rm in} < 10$ at 
0.008 $L_{\rm Edd}$.  If the value of $r_{\rm in} = 2$--3 measured
in the SPL state represents the location of the ISCO, then our hard 
state measurements imply a change of no more than a factor of 2.5 
and 5 (for the two luminosities, respectively) greater than the 
radius of the ISCO (although see caveats discussed above).

Recent theoretical work suggests a possible geometry that may 
be consistent with these observations.  It is found that two
physical processes can be important to causing material to 
condense out of an ADAF, leaving an inner optically thick disk.
First, as an ADAF is forming, the material close to the ADAF/disk
boundary will be significantly cooled via conductive cooling, 
and will cause ADAF material to recondense back into the disk
\citep{mlm07}.  Secondly, the soft photons from the optically
thick disk can Compton-cool the ADAF, which also leads to 
condensation \citep{liu07}.  For a relatively large range of
mass accretion rates and viscosity parameters, these effects
lead to inner and outer optically thick disks with an ADAF
filling the gap in between.  As discussed in \cite{liu07},
such a model would apply for the brighter portion of the hard
state and could explain the small inner radii inferred from
reflection fits in the hard state and the $\sim$0.2--0.3 keV
thermal components.  This geometry is also at least qualitatively 
consistent with the observed $\Omega/2\pi$ values.

Although such condensation can occur for relatively bright 
portions of the hard state, the calculations of \cite{liu07}
still indicate that below $L/L_{\rm Edd}\sim 0.001$, the inner
disk will evaporate, leaving only ADAF inside some truncation
radius.  Thus, it is notable that our observations of GX~339--4
as well as most, if not all, of the cases where small inner disk
radii have been inferred for the hard state via reflection
modeling or the presence of soft components have occurred above
this level \citep{miller06b,rykoff07}.  At lower luminosity, 
there have been observations that do not necessarily require 
soft components in the X-ray band.  For example, for the black hole 
system XTE~J1118+480, \cite{mcclintock01} find evidence for a soft 
component at the very low temperature of $\sim$24 eV when the 
system was near 0.001 $L_{\rm Edd}$ (assuming a distance of 
1.8 kpc and a black hole mass of 7\Msun).  In addition, 
\cite{tkk04} observed XTE~J1650--500 at levels of $10^{-4} 
L_{\rm Edd}$ (assuming a distance of 4 kpc and a black hole mass 
of 10\Msun), and did not detect a soft component or iron features.
Finally, very high quality spectra have been obtained for several
systems in quiescence $L/L_{\rm Edd}\sim 10^{-6}$ or lower 
without evidence for a thermal component or iron features 
\citep[e.g.,][]{bradley07,ctk06}.

\subsection{Implications for the Compact Jet}

The possibility of an inner optically thick disk in the hard state 
has very interesting implications for the production of compact
jets in the hard state.  For GX~339--4, we detect the compact jet
in the radio band contemporaneously with the times that we 
obtained Spectra \#1 and \#2, and \cite{miller06a} also report 
the presence of a compact jet during their hard state observation.  
Cygnus~X-1 provides another example of a bright hard state black 
hole with a compact jet.  These examples imply that a compact jet 
can be produced when the inner optically thick disk is present.  
Furthermore, it is notable that fitting multi-wavelength spectral
energy distributions from radio-to-X-ray observations of GX~339--4, 
Cygnus~X-1, and another black hole system, GRO~J1655--40, with
a compact jet model give values of 3.5--10 $R_{\rm g}$ for the 
radius of the jet at its base \citep{mnw05,migliari07}.  These
small radii are consistent with the jet being launched at
or within the inner edge of the disk, and may imply that the 
production of the compact jet is closely linked to the inner disk.

However, at the same time, radio observations show that compact 
jets can also be produced at very low luminosities by quiescent 
black holes \citep{gfh05,gallo06}.  This brings the role of the
inner disk in jet production into question since it is unclear
whether the inner disk persists to these low luminosities.  As
discussed above, the conductive and Compton cooling mechanisms 
considered by \cite{liu07} indicate that the disk should evaporate
below 0.001 $L_{\rm Edd}$.  However, it may still be worth 
considering whether other cooling mechanisms can maintain the
inner disk to lower levels.  For example, \cite{meier05} has 
explored the possibility of Magnetically Dominated Accretion
Flows (MDAFs), and there are indications for an inner disk region
in the hard state where magnetic cooling is important.  Clearly,
more theoretical studies as well as observations at these very
low luminosities are important for a full understanding the
full set of conditions that are required for compact jet production.

\clearpage

\acknowledgments

JAT would like to thank Sergio Campana for information about the
{\em Swift}/XRT calibration.  JAT thanks Tomaso Belloni, Jon Miller, 
and Jeroen Homan for useful discussions.  We would like to thank 
Neil Gehrels for approving the second set of {\em Swift} observations.
We appreciate comments from the anonymous referee that helped to
improve this paper.  JAT acknowledges partial support from NASA 
{\rm RXTE} Guest Observer grants NNG06GA81G and NNX06AG83G.  EK 
is supported by TUB\.ITAK Career Development Award 106T570 and also 
by a Turkish National Academy of Sciences Young and Successful 
Scientist Award.




\begin{table}
\caption{Observations of GX 339--4\label{tab:obs}}
\begin{minipage}{\linewidth}
\begin{tabular}{cccccc} \hline \hline
Satellite & Observation & Date in & Start Time & Stop Time & Exposure\footnote{This is the exposure time on the target (GX~339--4) obtained by the XRT instrument (for {\em Swift}) or the PCA instrument (for {\em RXTE}).}\\
          & ID (ObsID)  & 2007    & (UT hour)  & (UT hour) & (s)\\ \hline \hline
\multicolumn{6}{c}{Spectrum \#1}\\ \hline
{\em Swift} & 00030943001    & May 25  & 17.10 & 23.80 & 6150\\
{\em RXTE}  & 92704-04-02-00 & May 25  & 17.99 & 20.13 & 4976\\
{\em RXTE}  & 92704-04-02-02 & May 25  & 21.14 & 21.69 & 1808\\ \hline
\multicolumn{6}{c}{Spectrum \#2}\\ \hline
{\em Swift} & 00030943002    & June 10 & 16.70 & 21.70 & 4685\\
{\em Swift} & 00030943003    & June 12 &  1.29 & 21.98 & 5509\\
{\em Swift} & 00030943004    & June 14 &  1.01 & 22.38 & 3204\\
{\em RXTE}  & 92704-03-28-00 & June 10 &  4.43 &  5.63 & 2320\\
{\em RXTE}  & 92704-03-29-00 & June 11 &  8.44 &  9.51 & 3392\\
{\em RXTE}  & 94704-03-29-01 & June 12 &  3.27 &  3.64 & 1056\\
{\em RXTE}  & 94704-03-30-00 & June 13 &  6.08 &  6.64 & 1952\\
{\em RXTE}  & 94704-03-31-00 & June 14 & 15.29 & 16.28 & 2112\\ \hline
\end{tabular}
\end{minipage}
\end{table}

\begin{table}
\caption{GX 339--4 Spectral Fits with Basic Models \label{tab:basic}}
\begin{minipage}{\linewidth}
\footnotesize
\begin{tabular}{lcccccc} \hline \hline
Model\footnote{PL is a power-law model.  DBB is the \cite{mitsuda84} disk-blackbody model.  SM is the smeared iron edge model from \cite{ebisawa94}.  For these fits, we fixed the width of the smeared iron edge to 10 keV.} & $N_{\rm H}$\footnote{Errors on all parameter values are 90\% confidence ($\Delta\chi^{2} = 2.7$).} & $\Gamma$ & $N_{\rm PL}$\footnote{Units are photons~cm$^{-2}$~s$^{-1}$~keV$^{-1}$ at 1 keV.} & $kT_{\rm in}$ & $N_{\rm DBB}$ & $\chi^{2}/\nu$\\
      & ($10^{21}$ cm$^{-2}$) &   &    & (keV) &    &  \\ \hline \hline
\multicolumn{7}{c}{Spectrum \#1}\\ \hline
PL     &               $3.1\pm 0.1$ & $1.66\pm 0.01$ & $0.113\pm 0.002$ &  --   &   --    & 651.8/213\\
PL+DBB &               $8.5\pm 0.7$ & $1.69\pm 0.01$ & $0.125\pm 0.003$ & $0.178\pm 0.007$ & $76,000^{+39,000}_{-26,000}$ & 532.0/211\\
SM$\times$PL &         $2.6\pm 0.1$ & $1.58\pm 0.01$ & $0.104\pm 0.002$ &  --   &   --    & 444.7/211\\
SM$\times$(PL+DBB) &   $7.2\pm 0.7$ & $1.59\pm 0.02$ & $0.109\pm 0.003$ & $0.203^{+0.012}_{-0.010}$ & $23,000^{+15,000}_{-10,000}$ & 338.3/209\\ \hline
\multicolumn{7}{c}{Spectrum \#2}\\ \hline
PL     &               $4.9\pm 0.2$ & $1.61\pm 0.02$ & $0.0329\pm 0.0008$ &  --   &   --    & 321.6/213\\
PL+DBB &               $9.0\pm 1.1$ & $1.64\pm 0.02$ & $0.0367\pm 0.0013$ & $0.157\pm 0.011$ & $35,000^{+30,000}_{-18,000}$ & 280.1/211\\
SM$\times$PL &         $4.2\pm 0.2$ & $1.50\pm 0.02$ & $0.0294\pm 0.0008$ &  --   &   --    & 210.6/211\\
SM$\times$(PL+DBB) &   $7.0\pm 1.4$ & $1.52\pm 0.03$ & $0.0315\pm 0.0015$ & $0.184^{+0.023}_{-0.022}$ & $5,900^{+8,700}_{-4,300}$ & 200.2/209\\ \hline
\end{tabular}
\end{minipage}
\end{table}

\begin{table}
\caption{GX 339--4 Spectral Fits with Reflection\label{tab:reflection}}
\begin{minipage}{\linewidth}
\begin{tabular}{lccc} \hline \hline
Parameter\footnote{Errors on all parameter values are 90\% confidence ($\Delta\chi^{2} = 2.7$).} & Spectrum \#1 & Spectrum \#2 & 2004 Spectrum\footnote{These are the parameters obtained by \cite{miller06a} for the 2004 {\em XMM-Newton} and {\em RXTE} observations of GX~339--4.}\\ \hline \hline
$N_{\rm H}$ ($10^{21}$ cm$^{-2}$) & $7.1^{+0.9}_{-0.8}$ & $7.3^{+1.5}_{-1.3}$ & $3.7\pm 0.4$\\
$\Gamma$ & $1.68^{+0.03}_{-0.02}$ & $1.63^{+0.04}_{-0.03}$ & $1.41\pm 0.03$\\
$N_{\rm PL}$\footnote{Units are photons~cm$^{-2}$~s$^{-1}$~keV$^{-1}$ at 1 keV.} & $0.118^{+0.008}_{-0.006}$ & $0.035^{+0.004}_{-0.003}$ & $0.32\pm 0.03$\\
$kT_{\rm in}$ (keV) & $0.193\pm 0.012$ & $0.161^{+0.017}_{-0.031}$ & $0.39\pm 0.04$\\
$N_{\rm DBB}$ & $29,000^{+27,000}_{-14,000}$ & $14,000^{+35,000}_{-9,000}$ & $700\pm 200$\\
$i$ (degrees) & 20.0 & 20.0 & $20^{+5}_{-15}$\\
$\Omega/2\pi$\footnote{Reflection covering factor.} ({\ttfamily pexriv}) & $0.22^{+0.06}_{-0.05}$ & $0.24^{+0.11}_{-0.08}$ & $0.22\pm 0.06$\\
$\xi$\footnote{Ionization parameter.} (ergs cm$^{-1}$ s$^{-1}$) ({\ttfamily pexriv}) & $10,000^{+6,000}_{-5,000}$ & $7,000^{+8,000}_{-4,000}$  & 1,000\\
$E_{\rm line}$ (keV) & $6.9^{+0.2}_{-0.5}$ & $6.7^{+0.4}_{-0.3}$ & $6.8\pm 0.1$\\
$N_{\rm line}$ (photons cm$^{-2}$ s$^{-1}$) & $(7.4\pm 4.4)\times 10^{-4}$ & $(2.4^{+2.6}_{-2.1} )\times 10^{-4}$ & $(3.5\pm 0.3)\times 10^{-3}$\\
$R_{\rm in}/R_{\rm g}$ ({\ttfamily kdblur}) & $3.6^{+1.4}_{-1.0}$ & $2.9^{+2.1}_{-0.7}$ & $4.0\pm 0.5$\\
$q$\footnote{The power-law index for the radial emissivity profile.} ({\ttfamily kdblur}) & $3.2^{+0.5}_{-0.6}$ & $3.1\pm 0.4$ & $3.0$\\
$\chi^{2}/\nu$ & 324.7/205  & 191.3/205 & 2120.5/1160\\ \hline
Absorbed Flux (ergs cm$^{-2}$ s$^{-1}$)\footnote{Absorbed flux in the 1--100 keV band.} & $2.1\times 10^{-9}$ & $7.4\times 10^{-10}$ & $5.4\times 10^{-9}$\\
Unabs. Flux (ergs cm$^{-2}$ s$^{-1}$)\footnote{Unabsorbed flux in the 1--100 keV band.} & $2.2\times 10^{-9}$ & $7.7\times 10^{-10}$ & $5.5\times 10^{-9}$\\
Luminosity (ergs s$^{-1}$)\footnote{1--100 keV luminosity assuming a distance of 8 kpc.} & $1.7\times 10^{37}$ & $5.9\times 10^{36}$ & $4.2\times 10^{37}$\\
Luminosity ($L_{Edd}$)\footnote{1--100 keV luminosity in Eddington units assuming a distance of 8 kpc and a black hole mass of 5.8\Msun.} & 0.023 & 0.008 & 0.056\\ 
Iron Line Equivalent Width (eV) & $140\pm 90$ & $140^{+150}_{-120}$ & $\sim$160\\ \hline
\end{tabular}
\end{minipage}
\end{table}

\end{document}